\documentstyle[preprint,aps,psfig,pra]{revtex}
 \draft
\begin{document}
 \title{Existence of superposition solutions for pulse propagation in nonlinear resonant
 media}
 \author{P. K. Panigrahi \thanks{prasanta@prl.res.in} $^{1,2}$
 and G. S. Agarwal \thanks{gsa@prl.res.in} $^1$}
 \address{$^1$ Physical Research Laboratory,
 Navrangpura, Ahmedabad-380 009, India\\
 $^2$ School of Physics, University of Hyderabad, Hyderabad-500 046, India}
 %\date{\today}
 \maketitle

 \begin{abstract}
 Existence of self-similar, superposed pulse-train solutions of the nonlinear,
 coupled Maxwell-Schr\"odinger equations, with the frequencies controlled
 by the oscillator strengths of the transitions, is established.
 Some of these excitations are specific to the
 resonant media, with energy levels in the configurations of $\Lambda$ and $N$
 and arise because of the interference effects of cnoidal waves, as
 evidenced from some recently discovered identities involving the Jacobian
 elliptic functions. Interestingly, these excitations also admit a dual
 interpretation as single pulse-trains, with widely different
 amplitudes, which can lead to substantially different field intensities and
 population densities in different atomic levels.
 \end{abstract}
\vskip1cm PACS number(s): 42.50.Md, 42.50.Hz, 42.65.Tg

\newpage

\section{Introduction}

 The generation of shape preserving excitations in nonlinear media has been the
 subject of extensive research in diverse areas of physics, ranging from
 hydrodynamics \cite{Whitham}, particle physics \cite{Jackiw} to quantum
 optics \cite{Mac,Steudel,Grobe}
 and optical communications\cite{GPA}. Starting from the explanation of the
 solitary waves
 in shallow water\cite{kdv}, these  solutions of the nonlinear wave equations
 have found ample experimental verifications. In optical context, the nonlinear nature of
 the coupled Maxwell- Schr\"odinger equations \cite{Eberly1}, describing the interaction of
 classical radiation with matter in a resonant media, has naturally generated
 tremendous interest in the
 study of the  pulse \cite{Mac,Lamb1,Lamb2,Eberly2,Rahman,GSA}  and pulse-train
 solutions\cite{Eberly3,Bar,Hioe,Newbold}. Recently the continuous pulse-train soliton solutions
 have been observed experimentally for the two-level system \cite{Salamo}.
 For a detailed overview, the interested
 readers are referred to Refs.(\cite{maimi,And}), apart from the above
 references.\\
 The common factor that governs
 the existence of the self-similar excitations in various nonlinear systems is
 the critical balance between nonlinearity and dispersion \cite{Das}. This severely
 restricts the set of solutions, in most of the cases, to combinations of
 the Jacobian elliptic functions \cite{Jacobi} and also leads to
 definite relationships between the amplitudes, widths and velocities of these
 modes. The periodic elliptic functions, characterized by the modulus parameter
 $m$ and smoothly interpolating between hyperbolic
 secant and tangent to periodic cosine and sine functions, for $m$ values one
 and zero respectively, describe localized
 pulses for $m=1$ and pulse-trains for $0\leq m<1$.

 In this light,
 the recent findings of a number of identities involving the superposition
 of elliptic functions \cite{Sukh,Khare,Arul} and their application to various
 nonlinear systems \cite{Uday}
 makes it extremely interesting to find the implications of the same for
 the multi-level resonant media.
 A number of phenomena, like self \cite{Hahn} and electromagnetically induced
 transparencies \cite{Harris2}, pulse sharpening \cite{Bol}, pulse cloning and dragging
 \cite{Vemuri}, just to name a
 few, are ascribable to the existence of shape-preserving excitations in
 resonant media. A great deal of attention, both analytical and numerical, has been paid to
 the cases in which the resonant atoms are asymptotically in the ground or
 excited state. Pulse train solutions, characteristic to excited media, have
 been studied, for two-level \cite{Eberly3,Bar} and  for three and five-level \cite{Hioe}
 configurations. Well-known techniques for
 generating solutions e.g., inverse scattering methods \cite{Das,Bols,Bash} and B{\"a}cklund
 transformations \cite{Steu} have also been employed for higher level atomic media, albeit under
 restrictive conditions on the propagation constants and other parameters of these systems.
 Further progress, in the understanding of the
 dynamics of the inhomogeneously broadened three level system, has recently been achieved by the derivation
 of an area theorem \cite{Eberly}.

 In this paper, we first show the existence of novel
 pulse train solutions, specific to the $\Lambda$ and $N$-type, nonlinear,
 resonant media, taking recourse to the above mentioned identities, involving elliptic functions.
 Interestingly,
 these solutions can either be viewed as a linear superposition of cnoidal
 waves or as single cnoidal waves, with widely different, modulus dependent
 amplitudes for different pulse trains. We find that, for some of these
 exact solutions,
 the modulus parameter is controlled by the oscillator
 strengths of the atomic transitions, as compared to other pulse train solutions
 found so far, where $m$ appears as a free parameter.
 Interference effects, originating from the superposition nature of these solutions,
 lead to substantially different field intensities and population densities
 in different atomic levels. We then point out the existence
 of pulse train solutions in the four-level case, similar to the ones found in other
 multi-level systems.

 \section{Superposed pulse trains in $\Lambda$ systems}

 We start with the three-level $\Lambda$ system, because of its
 wide applicability and then proceed to the four-level case, with
 the energy levels in the configuration of $N$, from which the
 other lower ones will follow under limiting conditions.
Shape preserving solutions, in the form of
 superposition of a pair of {\it different} cnoidal waves (like a superposition
 of $\rm{sn}(x,m)$ and $\rm{cn}(x,m)$), have been obtained earlier by Hioe and
 Grobe \cite{Hioe}, for the three and five level systems. These exact solitary waves
can have a variety of
 shapes, because of interference.
 The solutions to be discussed here are superpositions of an odd number
 of cnoidal waves of a {\it given} type (no combinations involving, say both
 $\rm{sn}(x,m)$ and $\rm{cn}(x,m)$), with appropriately displaced arguments.
 As will be seen later in the text, these superposed solutions can have widely different
 amplitudes. This can
 be understood from the fact that, both $\rm{cn}(x,m)$ and  $\rm{sn}(x,m)$ functions take
values ranging from $-1$ to $1$, whereas the
 $\rm{dn}(x,m)$ function has only positive values. Hence, the superposed solutions involving
 $\rm{cn}(x,m)$ and  $\rm{sn}(x,m)$ functions can have a much smaller amplitude as compared to the one
 involving $\rm{dn}(x,m)$ function, whose amplitude can be substantially higher.

 For the $\Lambda$ system consisting of the ground state $i$, excited state $e$
 and the intermediate state $f$, we assume
 the fields to have slowly varying pulse envelopes:
 $\rm{\vec{E_{\alpha}}(z,t)}=\vec{\cal {E_{\alpha}}}(z,t)e^{-i(\omega_{\alpha}
  t-k_{\alpha} z)}~+~c.c.~.$ In the
  rotating wave approximation, ${\vec{\cal E}}_e$,
  and ${\vec{\cal E}}_f$ act on transitions $ei$ and $ef$, respectively.
  The Rabi  frequencies,
  $\Omega_{\alpha}=2\frac{\vec{d_{\alpha}} \vec{\cal{E_{\alpha}}}}{\hbar}$,
  with $\vec{d_{\alpha}}$ being the dipole matrix element of the
  $\alpha$-th transition, are also slowly varying functions of space and time.
  We assume, for simplicity that, all the fields are resonant with their
  respective transitions.

In slowly varying envelope approximations, the resonant coupled
Maxwell-Schr\"odinger equations are
 \begin{eqnarray}
 i\frac{\partial}{\partial \tau}C_{e}&=&-\frac{\Omega_f}{2}
 C_{f}-\frac{\Omega_e}{2}C_i~,\nonumber\\
 i\frac{\partial}{\partial \tau}C_{i}&=&-\frac{\Omega_{e}^{\star}}{2}
 C_{e}~,\nonumber\\
 i\frac{\partial}{\partial \tau}C_{f}&=&-\frac{\Omega_{f}^{\star}}{2}C_{e} \\
 \nonumber\\
 \frac{\partial}{\partial \zeta}\Omega_{f}&=&i\mu_{f}C_{e}C_{f}^{\star}~,\nonumber\\
 \frac{\partial}{\partial \zeta}\Omega_{e}&=&i\mu_{e}
 C_eC_{i}^{\star}~.
 \end{eqnarray}
 In above, we have neglected the relaxation terms, since the pulse widths are
 taken to be smaller than the relaxation times. $C_{\alpha}$
 ($\alpha=e,i,f)$ represents the probability amplitude of
 finding the atom in the state $|\alpha>$.

The parameter $\mu_{\alpha}$ is given by $\mu_\alpha=4\pi{\cal
N}\frac{|d_\alpha|^2 \omega_\alpha}{\hbar c}=\frac{2\pi e^2{\cal
N}f_\alpha}{mc}$ where $\cal {N}$ is the density of the atoms and
$f_{\alpha}$ being the oscillator strength for the transition of
frequency $\omega_{\alpha}$. The atomic system is assumed to be
continuously distributed in a non-dispersive host medium. The
coordinate $\tau=t-\frac{z}{c}$ measures time relative to the
pulse center and $\zeta=z$.

We look for self-similar solutions of Eqs. (1) and (2) i.e.,
solutions depending upon a single variable,
$X=(q\zeta-\Gamma\tau)$. Here, ${\Gamma}^{-1}$ is the pulse
duration and, as will be seen later, $q$ and $\Gamma$ will be
related via the pulse velocity and another parameter,
characterizing the solutions. The superposed pulse-train
solutions, to be discussed here, appear in both $\Lambda$ and $N$
type media and can be made to satisfy a wide range of initial
conditions, like no occupancy of the desired atomic levels to
partial occupation of all the levels. It should be pointed out that partial
occupation of levels necessarily implies initially induced coherence in the atomic system.
 We note that, the pulse train
solutions of $\Lambda$, $V$ and other odd-level systems
\cite{Hioe}, consisting of a matched pair of elliptic functions,
do not satisfy the nonlinear equations of the $N$ system.

It can be shown that, the
following ansatz solutions:
\begin{equation} \label{sup}
C_{i}=\tilde{S}~,C_{f}=b_{f}\tilde{C}~,C_{e}=b_{e}\tilde{D}~,
\Omega_{f}=A_{f}\tilde{S}~,\Omega_{e}=A_{e}\tilde{C}~;
\end{equation}
satisfy the Maxwell-Schr\"odinger equations, provided the
functions $\tilde{C},\tilde{S}$ and $\tilde{D}$ are as defined
below and the constant coefficients $b_f,b_e$ and $A_e,A_f$ are
appropriately related. $\tilde{C},\tilde{S}$ and $\tilde{D}$ are
linear superpositions of the Jacobi elliptic functions
$\rm{cn}(X,m), \rm{sn}(X,m)$ and $\rm{dn}(X,m)$,  respectively:
$\tilde{S}=\sum_{i=1}^p \rm{sn}[X+\frac{4(i-1)K(m)}{p},m]$ and
$\tilde{C}$ and $\tilde{D}$ are analogously defined. Here, $p$ is
an odd integer, $K(m)$ is the complete elliptic integral of the
first kind and $m$ is the modulus parameter. The above solutions
are possible for odd integral values of $p$, since the cross terms
in the right hand side of Eqs.(1) and (2) are cancelled because of
the identities of the type \cite{Sukh},
 \begin{eqnarray}
 \tilde{s_1}(\tilde{d_2}+\tilde{d_3})+{\mathrm cyclic~ permutations}=0~,\nonumber\\
 \tilde{c_1}(\tilde{s_2}+\tilde{s_3})+ {\mathrm cyclic~ permutations}=0~,\nonumber\\
 \tilde{c_1}(\tilde{d_2}+\tilde{d_3})+{\mathrm cyclic~ permutations}=0~.
 \end{eqnarray}
 Here, $\tilde {s_1}\equiv\rm{sn}(X,m)$ ,
 $\tilde {s_2}\equiv\rm{sn}(X+\frac{4K(m)}{p},m)$ and
 $\tilde {s_3}\equiv\rm{sn}(X+\frac{8K(m)}{p},m)$ and other functions are
 similarly defined.
 Although, for definiteness, henceforth we consider only the
 $p=3$ case, the
consistency conditions on the parameters, written below, are
identical for all values of $p$. Introducing an additional parameter $\mu=2q\Gamma m$, for future
convenience, one finds,
 \begin{eqnarray}
 |A_{e}|^2&=&4{\Gamma}^2m\frac{\mu_{e}}{\mu}~,\nonumber\\
 |A_{f}|^2&=&4{\Gamma}^2m[\frac{\mu_{e}}{\mu}-1]~, \label{1eq}\\
\nonumber\\
 b_{e}&=& \frac {iqA_{e}}{\mu_{e}} ~, \nonumber\\ \label{2eq}
 b_{f}&=&-\frac{{A}^{\star}_f}{{A_{e}}^{\star}}~;
 \end{eqnarray}
 and also the constraint, $\mu_{e}=\mu_{f}$. The positive definite character of the pulse intensity,
 requires that $\mu_{e}\geq\mu $~.

 The superposed character of the
 solutions, for higher values of $p$, leads to
 significant differences between
 different pulse-train and electron amplitudes, as will be explicated later.
 It also manifests in the conservation of probabilities for
 the electrons. For the $p=3$ case, $\sum_{\alpha}|C_{\alpha}|^2=1$ leads to,
 \begin{equation}\label{mueq}
\frac{\mu_{e}-\mu}{\mu_{e}}=\frac{\frac{1}{m}-1+\frac{4}{m}(\tilde{q}^2+\tilde{q})}
 {\frac{1}{m}-1+\frac{2}{m}[(\tilde {q}+1)^2-\frac{m}{(\tilde{q}+1)^2}]}~.
  \end{equation}
  Here, $\tilde{q}=\rm{dn}(\frac{2}{3}K(m),m)$ and satisfies,
  \begin{equation}
  \tilde{q}^4+2\tilde{q}^3-2(1-m)\tilde{q}-(1-m)=0~.
  \end{equation}
 The values of $\tilde{q}$ range from one to zero continuously, when $m$ varies
 between zero and one. It can be checked that, when, $m=1$, the
 conservation law is identically satisfied, making the above
 equation vacuous; in this case one obtains the pulse solutions
 of the $\Lambda$ system.
 For $m=0$, right hand side takes value one, which is ruled out on physical
 grounds. For other values of $m$, the above equation
 can be
 numerically solved to obtain the range of values of $\mu$ for a
 given $\mu_{e}$. It should be noted that, since
 $\mu_{e}\geq\mu $, the variable $\frac{\mu_{e}-\mu}{\mu_{e}}$ takes
 values between zero
 and one. It is found that for the
 superposed solutions to exist, the condition
  $0.59\lesssim\frac{\mu_{e}-\mu}{\mu_{e}}< 1$ should be satisfied.
  As depicted in Fig.1,
\begin{center}
\input epsf
\leavevmode{\epsfxsize=3in\epsfbox{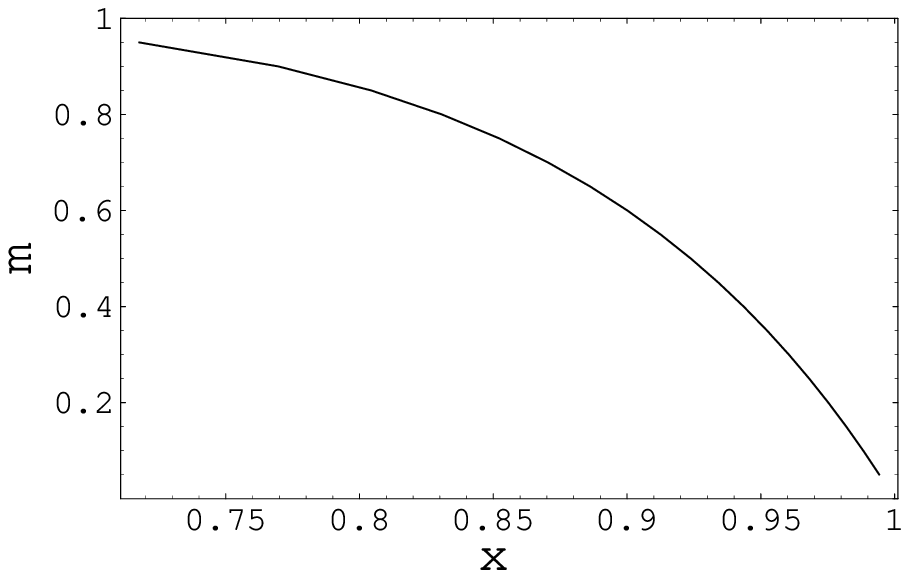}}
\end{center}
\begin{center}
FIG. 1. Graph of the modulus parameter $m$ versus $x=
\frac{\mu_e-\mu}{\mu_e}$.
\end{center}
 the modulus parameter $m$
 can lie between one and zero, for the above values of
 $\frac{\mu_{e}-\mu}{\mu_{e}}$. Hence the choice of $m$  $(0<m<1)$ determines $\mu$ through
 Eq.(\ref{mueq}), which in turn yields $q \Gamma$, through $\mu=2q \Gamma m$.
 Hence the solutions are characterized by the two free parameters $m$ and $ \Gamma$.
 It can be easily seen that
 ${\tilde D}$ has a much higher amplitude as compared to ${\tilde C}$
 and ${\tilde S}$; hence the $|e>$ state has a higher population.
 The superposition nature of the
 solutions enables one to have these widely differing amplitudes.
 It can be checked that exchange of
 ${\tilde C}$ and ${\tilde D}$ also leads to allowed solutions. In that case the
 $|f>$ state population can be
 made large as compared to $|i>$ and $|e>$ states.

 For the purpose of comparison, we consider the non-superposed $p=1$ case.
 Taking $C_{i}=b_{i}\rm{sn}(X,m)$ and
 replacing $\tilde{D}$ and $\tilde{C}$ by
 corresponding cnoidal waves one finds, after appropriately choosing the
 fields, that, the parameter
 relationships as obtained in Eqs.(\ref{1eq},\ref{2eq}) are unchanged.
  However, the
 probability conservation now yields substantially different result:
 \begin{equation} \label{9eq}
\frac{1}{|b_{i}|^2}=\frac{\mu}{\mu_{e}}(\frac{1}{m}-1)+1.
  \end{equation}
 It is clear from the above expression that, for, $b_{i}=1$, no pulse train
 solutions are
 possible, as $m=1$, for this case. Since $\mu / \mu_{e}$ lies between zero and
  one, it
 follows from Eq.(\ref{9eq}) that, $|b_{i}|^{2}$ takes values ranging from
 $m$ and $1$. It should be pointed that, like the previous case, $0<m<1$.
  One is then finally left with three free parameters,
  $\Gamma$, $m$ and $b_{i}$, in their appropriate range of values.
  It is possible to obtain another solution
 by exchanging $\rm{dn}(X,m)$ with $\rm{cn}(X,m)$, in the above
 ansatz. This choice leads to the constraint,
\begin{equation}
\frac{1}{|b_{i}|^2}=(1-\frac{\mu}{\mu_{e}})\frac{1}{m} \quad .
  \end{equation}
 \section{Pulse train solutions of the $N$ system}

 In case of the $N$ system, one
 needs to consider seven coupled equations, due to the presence
 of an additional energy state $|v>$, in between the states
 $|e>$ and $|f>$, of the $\Lambda$ system. Here, one has the additional parameter,
 $\mu_{v}= \frac{2 \pi e^2 {\cal N} f_v}{mc}$, $f_{v}$ being the corresponding oscillator strength.
 These nonlinear equations
 have been recently found to possess propagating pulse solutions \cite{GSA}.
 Introducing, the additional electron amplitude
 $C_{v}=b_{v}\tilde{D}$ and
 the Rabi frequency $\Omega_{v}=A_{v}\tilde{C}$ in the ansatz
 solutions, we only write below, the new and modified parametric relations, as
 compared to the $\Lambda$ system:
  \begin{eqnarray} \label{neq1}
  2q\Gamma m&=&\mu_{v}~,\\
  \nonumber\\
 |A_{v}|^2&=&4{\Gamma}^2m[1-\frac{\mu_{e}}{\mu_{f}}]~,\nonumber\\
 |A_{e}|^2&=&4{\Gamma}^2m\frac{{\mu_{e}}^2}{\mu_{f}{\mu_{v}}}~,\\
\nonumber \\
 b_{v}&=& \frac {iqA_{v}}{\mu_{v}} ~, \nonumber\\
 b_{f}&=&-\frac{A^{\star}_f}{{A}^{\star}_e}\frac{\mu_{e}}{\mu_{f}}~.
 \end{eqnarray}
 Note that here, $\mu_{v}$ determines $2q\Gamma m$, unlike the $\Lambda$ system.
 The positive definite character of the pulse intensities
 now requires the inequalities,
 \begin{equation}
 \mu_{f}\geq\mu_{e}\geq\mu_{v}~.
 \end{equation}

In the present case, the conservation of probability yields,
\begin{equation} \label{neq2}
\frac{\mu_{e}-\mu_{v}}{\mu_{f}}=\frac{\frac{1}{m}-1+\frac{4}{m}(\tilde{q}^2+\tilde{q})}
 {\frac{1}{m}-1+\frac{2}{m}[(\tilde {q}+1)^2-\frac{m}{(\tilde{q}+1)^2}]}~.
  \end{equation}
This leads to the constraint on the oscillator strengths:
 $0.59\lesssim\frac{\mu_{e}-\mu_{v}}{\mu_{f}}< 1$, for $0<m<1$. It is worth
 noting that, as compared to $\Lambda$-system, in the  present case the
 oscillator strengths determine modulus parameter $m$.

 The uniform group velocity, as determined by our
 solutions is given by,
 \begin{equation}
 \frac{1}{v}=\frac{1}{c}+\frac{\mu_{v}}{2\Gamma^2m}~.
 \end{equation}
 As the width of the pulse train becomes shorter ($\Gamma$ gets larger), the group
 velocity approaches the background medium velocity; however, for smaller values of
 $m$, the pulse train velocity can be significantly smaller than the medium
 velocity. We have checked that a different superposition type
 solution, e.g., $C_{i}=\tilde{S}$ and exchange of $\tilde{C}$ and $\tilde{D}$ for
 other fields, do not alter the above conclusions. However, the population of
 different levels, as well as the field amplitudes can be substantially different,
 from the earlier case. It is straightforward to see, from Eqs.(\ref{neq1}) and
 (\ref{neq2}), that unlike the $\Lambda$-system, the $N$-system is characterized
 by one free parameter $\Gamma$.

  It is worth noting that, for the two level case, corresponding to
  $\mu_{e}=\mu_{v}=\mu_{f}$, as well as for the $V$ system, for which
  $\mu_{e}=\mu_{v}$, the constraint equation originating from the probability
  conservation yields,
 \begin{equation}
 m=1+4(\tilde{q}^2+\tilde{q})~.
 \end{equation}
 Since $0\leq m \leq 1$ and ${\tilde q} > 0$, it is easy to see that, the
 above equation
 can not be satisfied. Hence these type of excitations are exclusive to
 $\Lambda$ and $N$ type systems.

A better understanding of
the above solutions can be provided by a dual interpretation of these excitations,
through a set of generalized Landen transformations, found recently. For the $p=3$ case, the
transformation formula yields,
\begin{equation}
\alpha\rm{dn}(x,\tilde{m})=[\rm{dn}(\alpha x,m) + \rm{dn}(\alpha x+\frac{4K(m)}{3},m)+
\rm{dn}(\alpha x+\frac{8K(m)}{3},m)]~,
\end{equation}
where $\alpha=[\rm{dn}(0,m) + \rm{dn}(\frac{4K(m)}{3},m)+\rm{dn}(\frac{8K(m)}{3},m)]$ and
\begin{equation}
\tilde{m}=m\frac{(1-\tilde{q})^2}{(1+\tilde{q})^2(1+2\tilde{q})^2}~,
\end{equation}
with $\tilde{q}=\rm{dn}(\frac{2}{3}K(m),m)$$\tilde {q}$.
 For the $\rm{sn}(x,\tilde{m})$ and $\rm{cn}(x,\tilde{m})$ functions, the formulae have same
 widths in right hand side as
 in $\rm{dn}(x,\tilde{m})$; however, the amplitude $\alpha$ is replaced by $\beta$, given by
 \begin{equation}
 \beta=[\rm{cn}(0,m) + \rm{cn}(\frac{4K(m)}{3},m)+
\rm{cn}(\frac{8K(m)}{3},m)]~.
\end{equation}
 The facts that, $\rm{dn}(x,m)$ takes only positive values and $\rm{cn}(x,m)$ have values ranging from
 $+1$ to $-1$, can be seen to be the reason behind the large value of $\alpha$ as compared to that of
 $\beta$. Hence, the superposed cnoidal waves can have widely varying amplitudes, which leads to
 significant differences in atomic level occupations.

 \section{Pure cnoidal solutions for the $N$ system}

 Since the four-level system has not been systematically analyzed for pulse
 train solutions, we would like to point out that, like the
 two-level case, pure cnoidal waves, with comparable electron and field
 amplitudes, can also be obtained as solutions to the $N$ system.
 As has been mentioned earlier, matched pair type pulse train solutions of the odd-level
 atomic systems \cite{Hioe} are not solutions of the four-level system. It
 can be seen that, the following ansatz solutions:
\begin{equation}
C_{i}=b_{i}\rm{sn}(X,m)~,C_{f}=b_{f}\rm{dn}(X,m)~,C_{e}=b_{e}\rm{cn}(X,m)~,
C_{v}=b_{v}\rm{dn}(X,m)~,\nonumber\\
\end{equation}
and
\begin{equation}
\Omega_{f}=A_{f}\rm{sn}(X,m)~,\Omega_{e}=A_{e}\rm{dn}(X,m)~,
\Omega_{v}= A_{v}\rm{dn}(X,m)~;\nonumber\\
\end{equation}
satisfy the Maxwell-Schr\"odinger equations, provided $|b_{i}|^2<1$ and the
constant coefficients $b_i,b_f,b_e,b_v$ and $A_e,A_f,A_v$ are related.
The consistency conditions on all the parameters are now modified:
 \begin{eqnarray}
 \frac{2q\Gamma m}{|b_{i}|^2}&=&\mu_{v}~,\nonumber\\
 |A_{v}|^2&=&4{\Gamma}^2[1-\frac{\mu_{e}}{\mu_{f}}]~,\nonumber\\
 |A_{f}|^2&=&4{\Gamma}^2m[\frac{\mu_{e}}{\mu_{v}}-1]~,\nonumber\\
 |A_{e}|^2&=&4{\Gamma}^2\frac{{\mu_{e}}^2}{\mu_{f}{\mu_{v}}}~,\nonumber\\
 b_{e}&=& \frac {i\mu_{v}A_{e}b_{i}}{2\mu_{e}\Gamma} ~, \nonumber\\
 b_{v}&=& \frac {iA_{v}b_{i}}{2\Gamma} ~, \nonumber\\
 b_{f}&=&-\frac{{A_{f}}^{\star}}{{A_{e}}^{\star}}
 \frac{\mu_{e}b_{i}}{m\mu_{f}}~,\nonumber\\ \label{4eq}
\frac{1}{|b_{i}|^2}&=&[1+\frac{1-m}{m}(\frac{\mu_{e}-\mu_{v}}{\mu_{f}})]~.
 \end{eqnarray}
For a given ground state occupancy, the modulus parameter $m$ is
determined by Eq.(\ref{4eq}).
 The velocity is now modified and depends upon $|b_{i}|^2$:
\begin{equation}
 \frac{1}{v}=\frac{1}{c}+\frac{\mu_{v}|b_{i}|^2}{2\Gamma^2m}~.
 \end{equation}
For the purpose of considering limiting cases of the $N$ system,
we observe that, when the modulus parameter $m$ equals one, the
cnoidal functions go over to hyperbolic ones, thereby yielding
localized pulse solutions. The pulse trains are obtained, when
$0<m<1$, the $m=0$ value being ruled out in the present case.
Since $\frac{\mu_{e}-\mu_{v}}{\mu_{f}}$ is positive, it is clear
from Eq. (\ref{4eq}) that, for the $N$ system, $|b_i|^2$ must be
less than one, for the latter type of solutions. However, for the
two level case, where $ \mu_{e}=\mu_{v}=\mu_{f}$, the above
restriction does not apply. In that case, $b_i=1$ and one obtains
the  cnoidal solutions, similar to  a set of solutions, obtained
by Crisp and Arecchi et al. in Ref. \cite{Eberly3}, in an
inhomogeneous media. Exchanging $\rm{dn}(X,m)$ with $\rm{cn}(X,m)$
in the above ansatz, leads to $|b_i|^2=m$; in this case one
obtains the solutions, analogous to another set of solutions given
in Ref. \cite{Eberly3}, under inhomogeneous conditions. This is
also clear from  the analysis of the $\Lambda$ system, which is
obtained from the $N$ system under the condition,
$\mu_{e}=\mu_{f}$, with $\mu =\mu_{v}$. The above point also
indicates that for the $\Lambda$ system, obtained as a limiting
case, $|b_i|^2<1$, in order for the pulse train solution to exist.
One obtains localized pulse solutions for $|b_i|^2=1$, since in
that case $m=1$. For the $V$ system, for which $ \mu_{e}=\mu_{v}$,
the conclusions are similar to the two-level case. A comparison
with Ref. \cite{Hioe} shows that the above solutions form a
subclass of the matched pair solutions, for the $\Lambda$ and $V$
system. This result is expected since matched pair type
superposition solutions are not possible for the $N$ system.

For the purpose of better appreciation of the relative amplitude
variations of the superposed solutions, as compared to the pure
cnoidal ones, Fig.2 depicts two of the pulse trains with their
cnoidal constituents. The widely different amplitudes of the superposed
${\rm dn}(X,m)$ and ${\rm sn}(X,m)$ type cnoidal waves and their respective
constituents are clearly visible.

\begin{center}
\input epsf
\leavevmode{\epsfxsize=3in\epsfbox{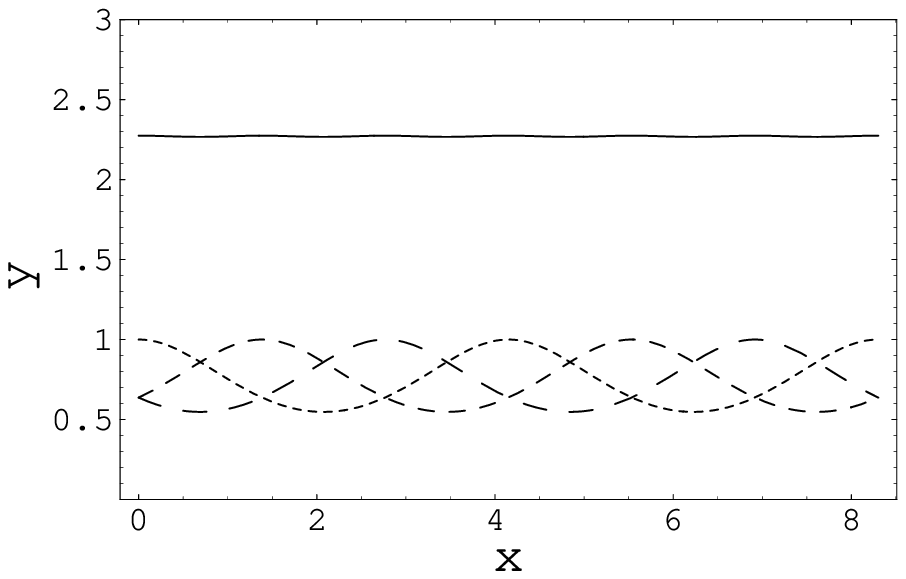}}
\end{center}
\begin{center}
FIG. 2a. Plot depicting the superposed ${\tilde D}(X,m)$ (solid
curve) and its three \\ ${\rm dn}(X,m)$ constituents (dotted
curves), for $m=0.7$.
\end{center}

\begin{center}
\input epsf
\leavevmode{\epsfxsize=3in\epsfbox{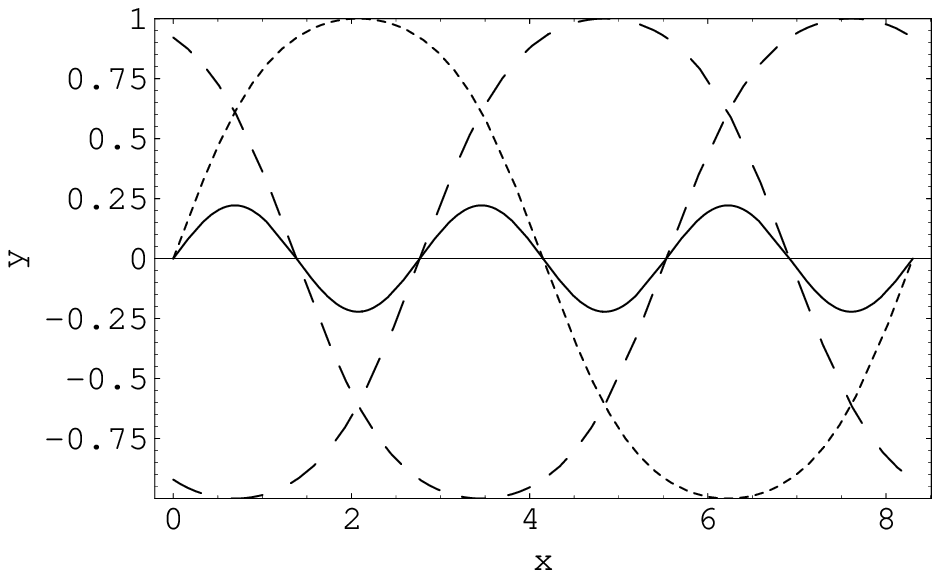}}
\end{center}
\begin{center}
FIG. 2b. Plot depicting the superposed ${\tilde S}(X,m)$ (solid curve)
and its three
\\ cnoidal constituents (dotted curve), for $m=0.7$.
\end{center}

 Fig.3 shows, the superposed solutions with their
non-superposed counterparts, clearly bringing out their
differences.

\begin{center}
\input epsf
\leavevmode{\epsfxsize=3in\epsfbox{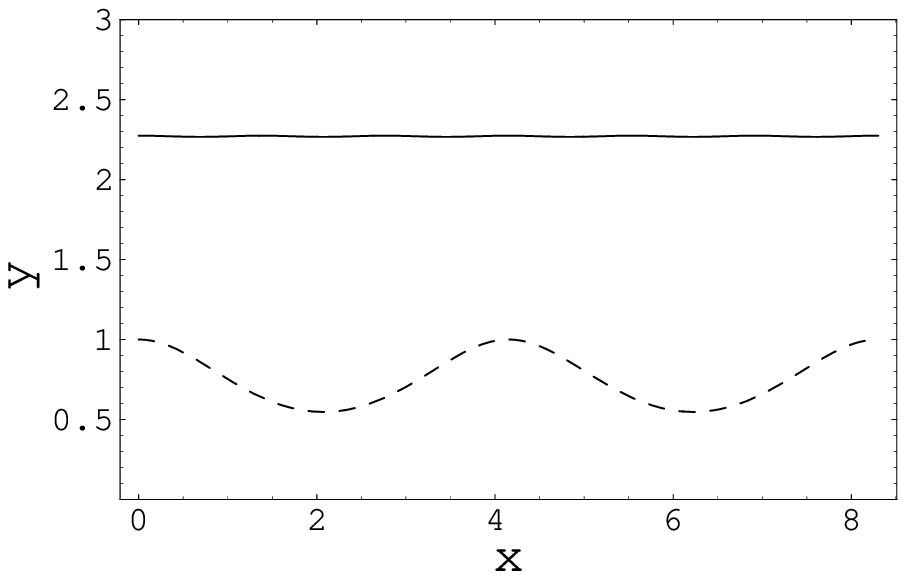}}
\end{center}
\begin{center}
FIG. 3a. Plots depicting the superposed solution ${\tilde D}(X,m)$ (solid
curve) of Eq.(\ref{sup}) and its non-superposed counterpart (dotted curve) in the same units, for
$m=0.7$. These plots represent the electron amplitudes in the state $|e> $ of the
$\Lambda$ system.
\end{center}

\begin{center}
\input epsf
\leavevmode{\epsfxsize=3in\epsfbox{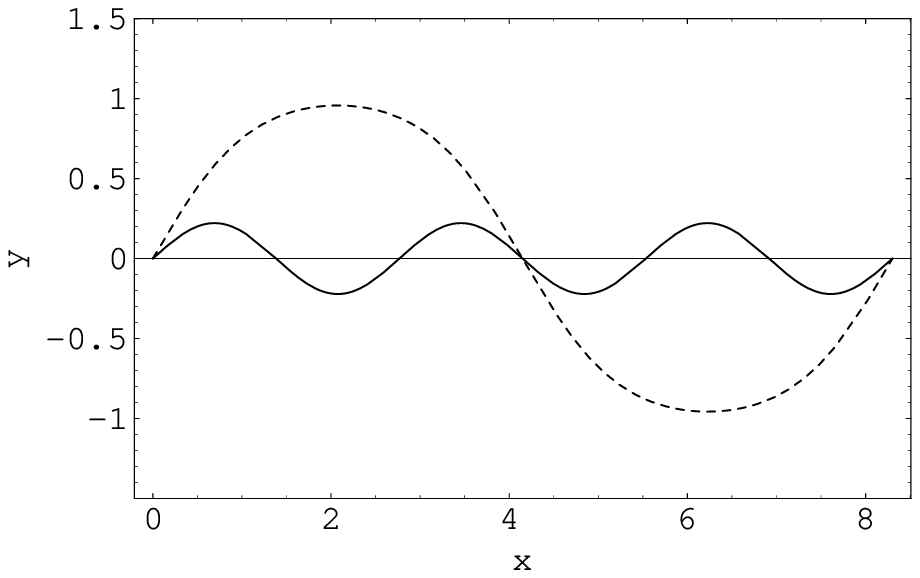}}
\end{center}
\begin{center}
FIG. 3b. Plots depicting the superposed $\rm sn(X,m)$ solution
 (solid curve) of Eq.(\ref{sup}) and its
 pure cnoidal counterpart (dotted curve), for $m=0.7$ and $b_{i}=0.9869$. With $b_{i}=1$ these
plots represent the field strengths $\Omega_{f}$ of the $\Lambda$
system, in the same units.
\end{center}

 Fig.4
reveals the same for different parameter values. One sees that,
for $m=0.3$ the amplitude of superposed ${\rm sn}(X,m)$ solution
attains a very small value.
 It is extremely interesting to observe
that, the excited state probability density can be made very small
and other densities enhanced, through these novel solutions, which
is not possible through pure cnoidal waves. It is clear that, this
desirable feature of the pulse trains owes its origin to the
superposition nature of the solutions.

\begin{center}
\input epsf
\leavevmode{\epsfxsize=3in\epsfbox{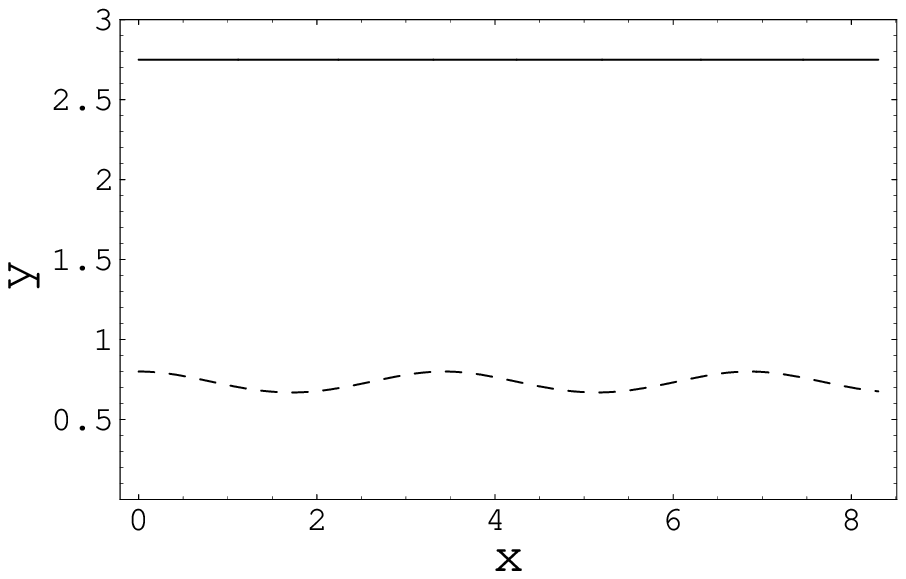}}
\end{center}
\begin{center}
FIG. 4a. Plots depicting the superposed ${\tilde D}(X,m)$ (solid
curve) and its non-superposed counterpart (dotted curve) in the same units, for
$m=0.3$.
\end{center}

\begin{center}
\input epsf
\leavevmode{\epsfxsize=3in\epsfbox{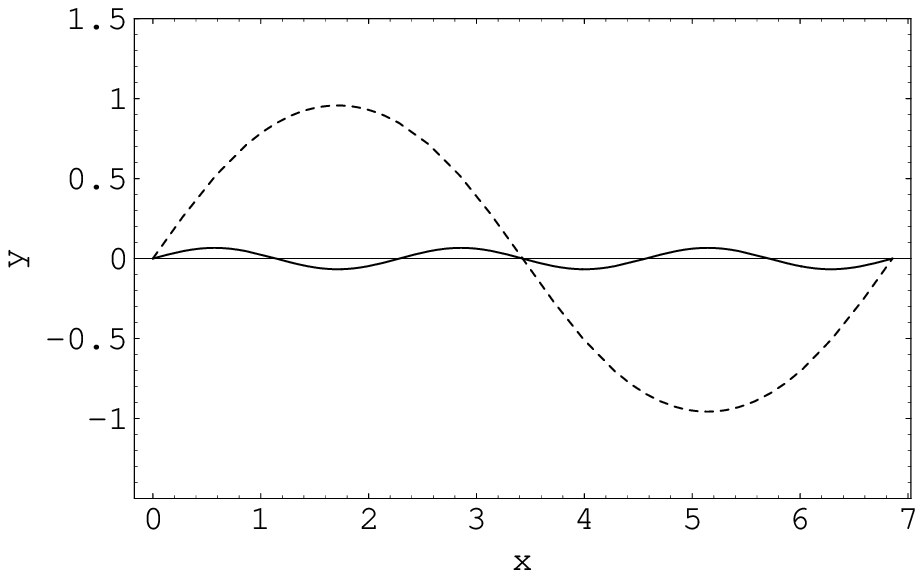}}
\end{center}
\begin{center}
FIG. 4b. Plots depicting the superposed $\rm sn(X,m)$ solution
 (solid curve) and its
 pure
\\ cnoidal counterpart (dotted curve), for $m=0.3$ and $b_{i}=0.9568$ .
\end{center}

It is natural to enquire the effect of detunings on the above solutions and to
find if the previous methods of incorporating  detuning (particularly the case
  $\Delta_{e}=\Delta_{f}$ in the $\Lambda$ system \cite{Rahman})
 would also work for deriving
  the pulse train solutions.
\section{Conclusions}

In summary, we have found that delicate interference phenomena
involving cnoidal waves give rise to certain novel
shape-preserving excitations, exclusive to $\Lambda$ and $N$ type
multi-level atomic media. It was also observed that $\it{matched~
pair}$-type self-similar solutions, possible in odd-level systems,
do not occur in the even-level $N$ system, although a variety of
simple pulse-train solutions are possible in the latter case.
Interestingly, the superposition nature of the above mentioned
excitations of the nonlinear media, makes it possible for
different atomic levels to have widely different population
densities and different pulse trains possess significantly
differing amplitudes, features not present in simple pulse-train
solutions. For the $N$-system, the frequencies of the propagating
pulses, which govern the spatio-temporal behavior of the atomic
population densities, are not arbitrary, as have been the case so
far, with the pulse-train solutions of various systems. The
modulus parameter $m$, which appears in the velocities and also
controls the period $K(m)$ is determined by the propagation
constants. In light of the potential usefulness of the multi-level
systems, starting from information storage \cite{Liu,Tarak} to
quantum computation \cite{Tich}, we hope that, some of these
desirable features of the excitations may find practical
applications. It is interesting to observe that, presence of
additional non-linearities leads to chirping of the pulse train of
the two-level systems \cite{Matu}. The effect of similar
non-linearities on the superposed and non-superposed pulse trains
solutions, for the three and four-level systems is
worth further investigation.  \\

One of the authors (PKP) acknowledges useful discussions with
Prof. A. Khare.

%\begin{thebibliography}{99}

%\end{enumerate}


\begin{references}

%\begin{center}
%{\large REFERENCES}
%\end{center}

%\begin{enumerate}

\bibitem{Whitham} C.B. Whitham, {\it Linear and Nonlinear Waves} (Wiley, New York, 1974).

\bibitem{Jackiw} R. Jackiw, Rev. Mod. Phys. {\bf 49}, 681 (1977).

\bibitem{Mac} S.L. McCall and E.L. Hahn, Phys. Rev. Lett. {\bf 18}, 908 (1967).

\bibitem{Steudel} H. Steudel, Physica {\bf 6D}, 155 (1983).

\bibitem{Grobe} R. Grobe, F.T. Hioe and J.H. Eberly, Phys. Rev. Lett. {\bf 73}, 3183 (1994).

\bibitem{GPA} G.P. Agrawal, {\it Nonlinear Fiber Optics} (Academic Press, New York, 1989);
A. Hasegawa, {\it Optical Solitons in Fibers} (Springer-Verlag, Berlin, 1989).

\bibitem{kdv} D.J. Korteweg and G. de Vries, Phil. Mag.(5) {\bf 39}, 422 (1895).

\bibitem{Eberly1} L. Allen and J.H. Eberly, {\it Optical Resonances and Two-Level Atoms} (Dover, New York,
1987).

\bibitem{Lamb1} G.L. Lamb, Jr., Phys. Lett. {\bf A 25}, 181 (1967).

\bibitem{Lamb2} G.L. Lamb, Jr., Rev. Mod. Phys. {\bf 43}, 99 (1971).

\bibitem{Eberly2} M.J. Konopnicki and J.H. Eberly,
Phys. Rev. {\bf A 24}, 2567 (1981).

\bibitem{Rahman} A. Rahman and J.H. Eberly, Phys. Rev. {\bf A 58}, R805 (1998).

\bibitem{GSA} G.S. Agarwal and J.H. Eberly, Phys. Rev. {\bf A 61}, 13404 (1999).

\bibitem{Eberly3} J.H. Eberly, Phys. Rev. Lett. {\bf 22}, 760 (1969);
F.T. Arecchi, V. DeGiorgio and S.G. Someda,
Phys. Lett. {\bf A 27}, 588 (1968); M.D. Crisp,
Phys. Rev. Lett. {\bf 22}, 820 (1969).

\bibitem{Bar} T.W. Barnard, Phys. Rev. {\bf A 7}, 373 (1973).

\bibitem{Hioe} F.T. Hioe and R. Grobe, Phys. Rev. Lett. {\bf 73}, 2559 (1994).

\bibitem{Newbold} M.A. Newbold and G.J. Salamo, Phys. Rev. Lett. {\bf 42}, 887
(1979).

\bibitem{Salamo} J.L. Shultz and G.J. Salamo Phys. Rev. Lett. {\bf 78}, 855
(1997).

\bibitem{maimi} A.I. Maimistov, A.M. Basharov, S.O. Elyutin and Yu.M.
Sklyarov,  Phys. Rep. {\bf{C}} {\bf 191}, 1 (1990).

\bibitem{And} A.V. Andreev, JETP {\bf 86}, 412 (1998).

\bibitem{Das} A. Das, {\it Integrable Models} (World Scientific, Singapore, 1989);
P.G. Drazin and R.S. Johnson, {\it Solitons: An Introduction}
(Cambridge Univ. Press, 1989).

\bibitem{Jacobi} H. Hancock, {\it Theory of Elliptic Functions} (Dover, New York, 1958);
M. Abramowitz and I. Stegun, {\it Handbook of Mathematical
Functions} (NBS, US Government Printing Office, 1964).

\bibitem{Sukh} A. Khare and U. Sukhatme, Jour. Math. Phys. {\bf 43}, 3798 (2002).

\bibitem{Khare} A. Khare and U. Sukhatme, math-ph/0204054.

\bibitem{Arul} A. Khare, A. Lakshminarayan and U. Sukhatme, math-ph/0207019.

\bibitem{Uday} A. Khare and U. Sukhatme, Phys. Rev. Lett. {\bf 88}, 244101 (2002).

\bibitem{Hahn} S.L. McCall and E.L. Hahn, Phys. Rev. {\bf 183}, 457 (1969).

\bibitem{Harris2} S.E. Harris, Phys. Today {\bf 50}(7), 36 (1997).

\bibitem{Bol} L.A. Bol'shov, N.N. Elkin, V.V. Likhanskii and M.I. Persiantsev, Sov. Phys. JETP {\bf
61}(1), 27 (1985).

\bibitem{Vemuri} G. Vemuri, G.S. Agarwal, Phys. Rev. Lett. {\bf 79}, 3889 (1997).

\bibitem{Bols} L.A. Bol'shov, N.N. Elkin, V.V. Likhanskii and P.A. Napartovich, Sov. J. Quantum Electron. {\bf
12}, 941 (1982); L.A. Bol'shov and V.V. Likhanskii, {\it ibid}. {\bf 15}, 889 (1985).

\bibitem{Bash} A.M. Basharov and A.I. Maimistov, Sov. Phys. JETP {\bf 67}, 2426 (1988).

\bibitem{Steu} H. Steudel, J. Mod. Opt. {\bf 35}, 693 (1988); Q.-H. Park and H.J. Shin, Phys. Rev. {\bf A
57}, 4643 (1998).

\bibitem{Eberly}J.H. Eberly and V.V. Kozlov, Phys. Rev. Lett. {\bf 88},
243604 (2002).


\bibitem{Liu} C. Liu, Z. Dutton, C.H. Behroozi, and L.V. Hau,
Nature (London) {\bf 409}, 490 (2001); D.E. Phillips, A. Fleischhauer, A. Mair,
R. L. Walsworth and
M.D. Lukin, Phys. Rev. Lett. {\bf 86}, 783 (2001).

\bibitem{Tarak} T.N. Dey and G.S. Agarwal, quant-ph/0209099.


\bibitem{Tich}W. Teich and G. Mahler, {\it Complexity, Entropy and
the Physics of Information}, SFI studies
in Sciences of Complexity, Vol. VIII, W. Zurek (ed.),
Addison-Wesley, Reading, MA, 289 (1990).

\bibitem{Matu} L. Matulic, P.W. Milonni and J.H. Eberly, Opt. Commun.
{\bf 4}, 181 (1971);
L. Matulic and J.H. Eberly, Phys. Rev. A {\bf 6}, 822, 1258 (1972).

%\end{thebibliography}

\end{references}
\end{document}